\documentclass[%
 reprint,
 amsmath,amssymb,
 aps,
]{revtex4-2}

\usepackage{graphicx}
\usepackage{dcolumn}
\usepackage{bm}

\usepackage{siunitx}
\usepackage[dvipsnames]{xcolor}

\def\K{$^{39}$K}
\def\Na{$^{23}$Na}
\def\NaK{\Na\K}

\newcommand{\ket}[1]{\left|#1\right\rangle}

\begin{document}

\preprint{APS/123-QED}

\title{Electric-field control of atom-molecule Feshbach resonances}

\author{Mara Meyer zum Alten Borgloh}
\email[]{mara.meyer@iqo.uni-hannover.de}

\author{Jule Heier}
\author{Fritz von Gierke}
\author{Baraa Shammout}
\author{Eberhard Tiemann}
\author{Leon Karpa}
\author{Silke Ospelkaus}
\affiliation{Institut für Quantenoptik, Leibniz Universität Hannover, Hannover, Germany}

\date{\today}

\begin{abstract}
Ultracold molecules provide opportunities for exploring quantum matter, chemical dynamics and information processing thanks to their rich interactions, which can be controlled by external fields. Magnetic fields tune interactions through Feshbach resonances, enabling the formation of ultracold dimers and triatomic molecules from atom–dimer collisions. Here we demonstrate electric-field control of atom–molecule Feshbach resonances. In mixtures of ground-state sodium–potassium molecules and potassium atoms, electric fields shift resonance positions systematically, revealing specific trimer bound states and their electric-field dependent energies. The response differs markedly from isolated dimers, showing hindered rotation of the molecular constituent near an atom. Electric fields therefore add an independent knob for atom-molecule resonances, open spectroscopic access to triatomic quantum states, and advance controlled polyatomic quantum matter.

\end{abstract}

\maketitle


\section{\label{sec:intro}Introduction}
Ultracold polar molecules extend the boundaries of quantum science beyond the realm accessible to atomic systems. Their long-range and an\-iso\-tropic interactions, together with rich internal degrees of freedom will widen the scope for exploring many-body phases, steering chemical reactions at ultralow temperatures, and implementing dipolar platforms for quantum information processing \cite{Cornish2024,Bohn2017}. Unlocking this potential requires precise control over molecular interactions, both molecule-molecule and atom-molecule. Major progress has been made toward this goal: collisional shielding techniques \cite{MWshieldProposalKarman,MWshieldProposalQuemener,EfieldShielding,Karam2023,Anderegg2021,ShieldingWill,ShieldingWang,FermigasMunich} have enabled the creation of degenerate gases of polar molecules \cite{FermigasMunich,BECwill,BECWangArxiv}, while resonance-based association and field-linked states have provided demonstrations of ultracold polyatomic species \cite{TrimersPan,FieldLinkedTetramerLuo}. These milestones highlight how new tools for controlling interactions can open fundamentally new regimes of molecular quantum science.

A central tool for controlling interactions in ultracold atomic systems has been the use of magnetic fields to access Feshbach resonances, where a weakly bound state couples to the scattering continuum and thereby tunes collisional properties. This method has enabled the creation of ultracold diatomic molecules \cite{FeshMole} and has recently been extended to atom-molecule mixtures \cite{ControllingCollisionsKetterle,AbInitioKarman,TrimerFeshisKetterle,Yang2019,Wang2021,Morita2024,Hermsmeier2021,RbandCaFMike2023}, producing triatomic species \cite{TrimersPan}. Polar molecules possess permanent electric dipole moments, suggesting that electric fields could provide a complementary means of tuning resonances. Such control would couple directly to the dipolar molecular structure, opening new opportunities for manipulating interactions and offering valuable benchmarks for theoretical models of triatomic resonances, which remain challenging to describe. Yet despite this promise, electric-field control of scattering resonances has been realized only in solid-state systems \cite{Schwartz2021} and has, until now, remained out of reach in ultracold gases.

Here we overcome this challenge by demonstrating electric-field control of magnetic Feshbach resonances in an ultracold atom–molecule mixture. We study ground-state \NaK\ molecules colliding with \K\ atoms prepared in different hyperfine states, and observe that applied electric fields systematically shift resonance positions. These shifts allow us to assign the resonances to specific trimer bound states and to extract their Stark shifts. The response deviates strongly from that of isolated dimers, pointing to hindered molecular rotation in the presence of an atom. Our findings extend the understanding of polyatomic Feshbach molecules and their control to a new degree of freedom, substantially widening the scope of studies and applications in these systems.

\section{\label{sec:results}Electric tuning of Feshbach resonances}

We start with an ensemble of $\num{2.0 \pm 0.5 e5}$ potassium atoms, prepared in a hyperfine state $\ket{F,\,m_F}_{\textrm{K,s}}$, where $F$ denotes the total angular momentum and $m_F$ its projection onto the magnetic field axis. The index s indicates the scattering partner whereas later the index b labels the bound potassium.  Alongside, we prepare $\num{8 \pm 2 e3}{}$ NaK molecules in their rovibronic ground state and in the hyperfine state $\ket{m_{I,\mathrm{Na}}=-3/2, m_{I,\mathrm{K}}=-1/2}$, where $m_I$ is the projection of the nuclear spin of Na, respectively K, onto the magnetic field axis. Both ensembles are held in a common crossed optical dipole trap at a temperature of $T=\SI{520\pm 50}{nK}$. Details of the preparation procedure are provided in Appendix \ref{app:ExpDetails} and in previous work \cite{Kai}.

\begin{figure}
    \centering
    \includegraphics[width=1\linewidth]{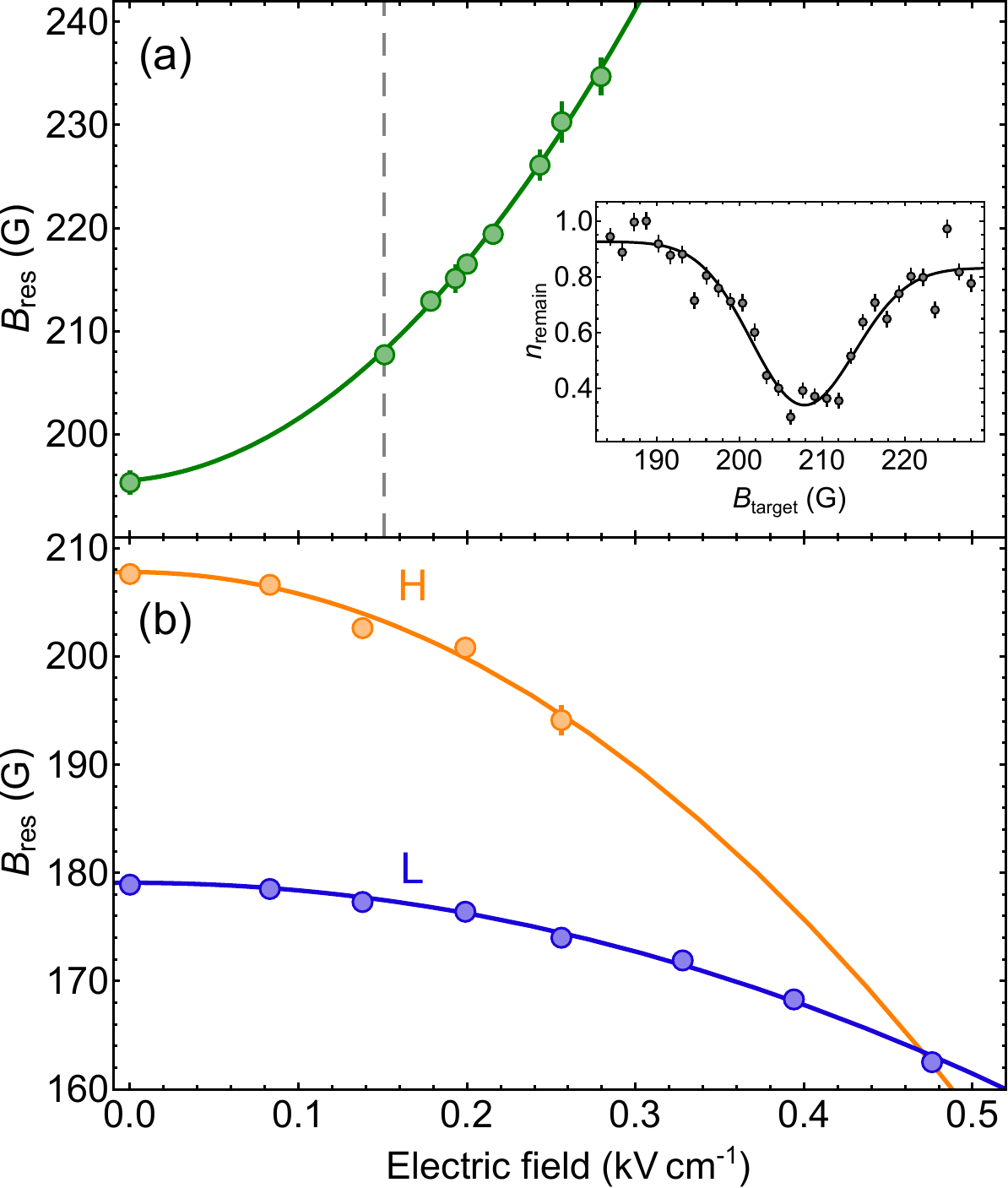}
    \caption{Electric-field dependence of magnetic resonance positions.
    Resonance positions $B_{\textrm{res}}$ are plotted versus the applied electric field $E$ for the collision partner K prepared in (a) $\ket{F=1, m_F=0}_{\textrm{K,s}}$ and in (b) $\ket{F=2, m_F=-2}_{\textrm{K,s}}$. Panel (a) shows a single resonance; panel (b) displays two distinct resonances labeled H (orange) and L (blue). Solid curves are quadratic functions of $E$ and serve as visual guides. Resonance positions are obtained from fits to molecule loss spectra versus $B_{\textrm{target}}$. An example spectrum for $E = \SI{0.15}{kV\,cm^{-1}}$ (dashed vertical line) is shown in the inset. Error bars in the inset represent the standard error of the relative molecule number from averaged images over 6-8 runs. To account for losses during the magnetic field ramp over the resonance, we use a Gaussian scaled with an error function as a fit model (inset, black solid line). Error bars (smaller than plot marker if not visible) in (a) and (b) correspond to the standard error from these fits. 
    \label{fig:EfieldData}} 
\end{figure}

We apply a fixed homogeneous electric field $E$ parallel to the magnetic field $B$ throughout the whole sequence. After preparing the mixture, $B$ is ramped to $B_{\textrm{target}}$, held for $\SI{10}{ms}$ to $\SI{15}{ms}$ and then returned to its initial setting for atom removal and molecular imaging. 
When varying $B$, we observe resonant losses around a characteristic magnetic field $B_\textrm{res}$ stemming from collisions of the dimers with the potassium atoms, as shown in the inset of Fig. \ref{fig:EfieldData}. Similar loss features, attributed to magnetic triatomic Feshbach resonances, have recently been reported in other atom–dimer mixtures \cite{Yang2019,Wang2021, ControllingCollisionsKetterle, TrimerFeshisKetterle}.

Expanding beyond previous studies, we make use of the capability to apply a precisely controlled external electric field to study its impact on these resonances. To this end, we measure $B_\textrm{res}$ for different electric field strengths $E$ as shown in Fig. \ref{fig:EfieldData}. For three resonances in the depicted states, we observe a clear dependence of $B_\textrm{res}$ on $E$. This establishes electric fields as a practical tool for controlling atom–molecule resonances due to the electric dipole moment of the molecular component, complementing the magnetic tuning that has so far dominated the field.

\section*{\label{sec:electrshift}Determining the triatomic bound state}

\begin{figure}[t]
    \centering
    \includegraphics[width=1\linewidth]{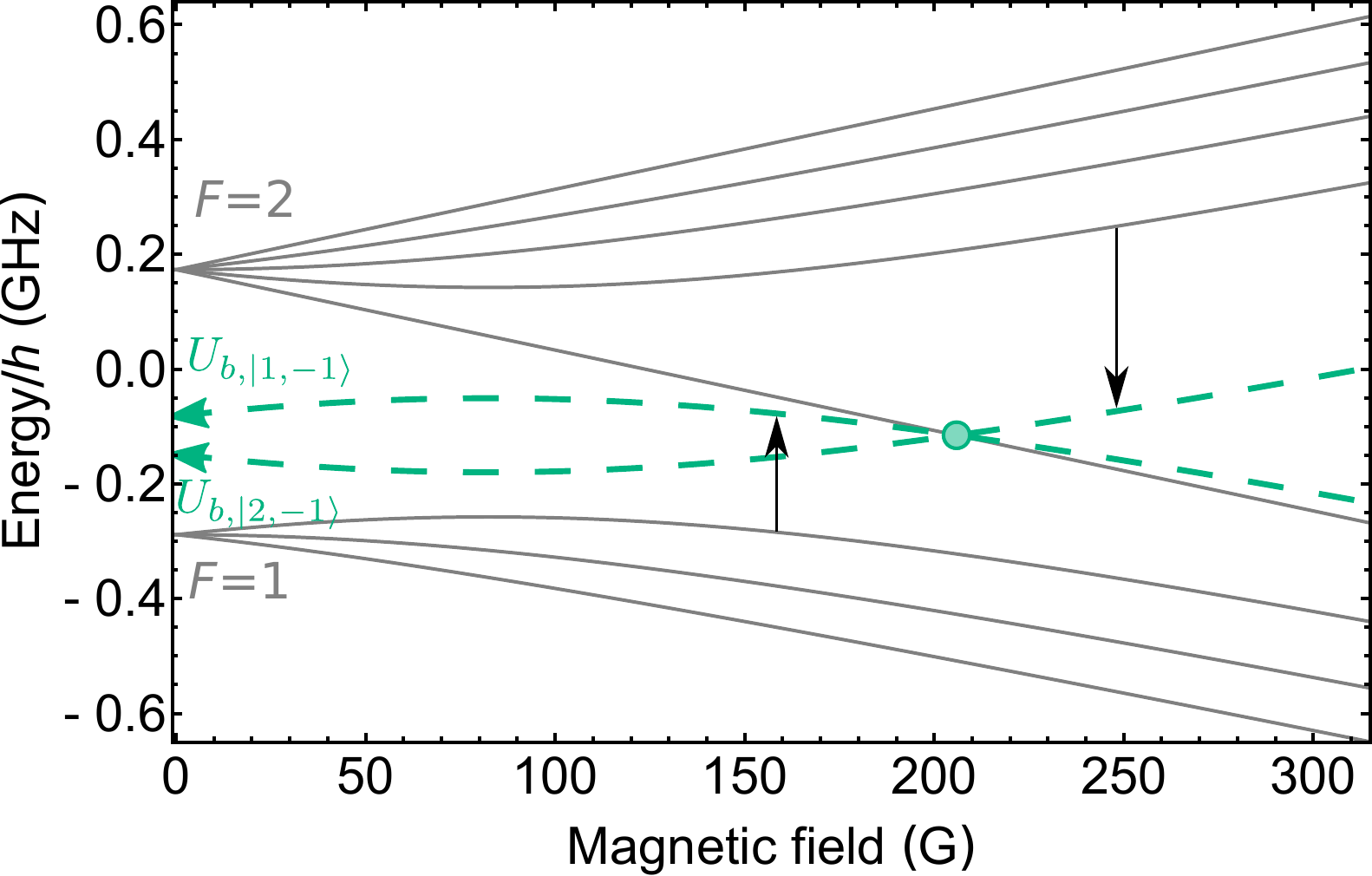}
    \caption{Method to determine the bound state energy. The energy of the scattering state as a function of magnetic field is approximated by the Zeeman effect of \K\ (gray solid lines). Energies are referenced to the center of gravity of the K hyperfine structure plus the Stark shift of the uncoupled NaK molecule. A measured resonance in the channel $\ket{F=2, m_F=-2}_{\textrm{K,s}}$ is shown as a green circle. By the selection rules for $\Delta m_F$ (see main text), this resonance can only originate from a bound state with $\ket{F=1, m_F=-1}_{\textrm{K,b}}$ or $\ket{F=2, m_F=-1}_{\textrm{K,b}}$. Propagating these states back to $B=0$ (green dashed lines), starting from the measured resonance, we can extract the possible energy of the bound state $U^{\textrm b}_{F,m_F}(E)$ (arrows) referenced to the Stark shift of the uncoupled NaK molecule for the applied electric field $E$. Knowing the energy of the scattering state, we obtain the energy of the trimer $U^{\textrm{NaK}_2}_{F,m_F}(E)=$$U^{\textrm{NaK}}_{N=0,m_N=0}(E)+U^{\textrm{b}}_{F,m_F}(E)$, including the Stark shift of NaK and referenced to zero fields and center of gravity of the hyperfine structure of K and the prepared state of NaK.} 
    \label{fig:DetermineEb}
\end{figure}

To uncover the nature of the triatomic bound states underlying the observed resonances, we analyze how their energies shift with the applied electric field $E$. 
Since the resonance occurs when the bound-state energy crosses the scattering threshold, any electric-field-induced differential shift between these two energies must be compensated by a change in magnetic field, leading to an electric-field-dependent resonance position $B_\textrm{res}(E)$.
A significant variation of $B_\textrm{res}$ therefore indicates that the electric response of the triatomic bound state differs from that of the incoming scattering state.
This provides a spectroscopic handle: the response of a resonance with $E$ reveals the Stark shift of the bound state and yields information about the internal structure of the trimer.  To quantify this behavior, we extrapolate the energy of the crossing trimer state $U^{\textrm{NaK}_2}_{F,m_F}(E)$ at $B=0\,\textrm{G}$, as described in Fig. \ref{fig:DetermineEb}. 
In the electronic ground state the molecule’s magnetic response is strongly suppressed, so the magnetic-field dependence of the threshold is dominated by the atomic Zeeman shift. As argued by Wang \textit{et al.} \cite{Wang2021}, the trimer considered here can be treated as a weakly bound state (see also \cite{Shammout2023}), with its magnetic properties dominated by potassium. In this picture, the angular-momentum projection of K follows a simple selection rule \cite{SelectionRulesFrye}: it may change only by $\Delta m_F=m_{F,\textrm{K,b}}-m_{F,\textrm{K,s}}=0,\pm 1$ between the scattering state and the bound state, with $m_{F,\textrm{K,b}}$ and $m_{F,\textrm{K,s}}$ being the projections of the potassium constituent for the trimer-bound and the scattering state, respectively.
A further constraint comes from the nature of magnetic Feshbach resonances, which require a finite difference in magnetic moment between the two states. Thus, scattering and bound states must differ at least in $F$ or $m_F$. With this restriction, we calculate $U^{\textrm{NaK}_2}_{F,m_F}$ as a function of the electric field for all possible trimer states that could give rise to the observed resonances, as outlined in Fig. \ref{fig:DetermineEb}.

These energies are fitted using
\begin{eqnarray}
U^{\textrm{NaK}_2}_{F,m_F}(E)=
{(\mu ^{\textrm{rel}}_{F,m_F, N, m_N}})^2 \, U^{\textrm{NaK}}_{N,m_N}(E&)\nonumber\\
+U^{\textrm {b}}_{F,m_F}(E=0&),
\label{eq:Fitformular}
\end{eqnarray}

\noindent where $F$ and $m_F$ denote the hyperfine state of the bound potassium atom, $N$ denotes the rotational angular momentum of the uncoupled dimer, $m_N$ its projection onto the quantization axis, and $U^{\textrm{NaK}}_{N,m_N}$ is the shift of the dimer state $\ket{N,m_N}$ due to electric field. The ratio $({\mu ^{\textrm{rel}}_{F,m_F, N, m_N}})^2$ describing the Stark shift of the trimer in relation to the state $\ket{N,m_N}_{\textrm{NaK}}$ and the energy of the field-free trimer $U^{\textrm {b}}_{F,m_F}(E=0)$ serve as fit parameters. Physically, $({\mu ^{\textrm{rel}}_{F,m_F, N, m_N}})^2$ quantifies how strongly the trimer follows the diatom’s Stark response for a given rotational component, i.e., an effective weighting of the diatomic character in the bound state.
The values of all fit parameters are presented in Table \ref{tab:FitPar} in Appendix \ref{app:DetailedDiscussion}.
By comparing the Stark effect of possible trimer states  with that of the free diatom and by taking into account the energy of the bound state in the field-free case, we derive constraints that allow for plausible assignment of the observed resonances. We exclude candidates that would require relative Stark shifts far exceeding the isolated-dimer response, and we require the resulting zero-field energies to remain bound with respect to the appropriate atom-molecule asymptote. The remaining candidates yield the assignments stated below. To the resonance observed in $\ket{1,0}_{\textrm{K,s}}$, we obtain $\ket{N=1,m_N=-1}_{\textrm{NaK}}\ket{F=1,m_F=1}_{\textrm{K}}$  with a relative Stark shift of $({\mu ^{\textrm{rel}}_{1,1, 1, \pm 1}})^2 = 1.11~(\pm 0.07)$. For the resonance H in $\ket{2,-2}_{\textrm{K,s}}$ we assign
$\ket{N=1,m_N=-1}_{\mathrm{NaK}}\ket{F=1,m_F=-1}_{\mathrm{K}}$, with $({\mu ^{\textrm{rel}}_{1,-1, 1, \pm 1}})^2 = 0.5~(\pm 0.4)$ and for L $\ket{N=1,m_N=0}_{\textrm{NaK}}\ket{F=2,m_F=-1}_{\textrm{K}}$ with $({\mu ^{\textrm{rel}}_{2,-1, 1, 0}})^2 = 0.51~(\pm 0.05)$.

\begin{figure*}[t]
\includegraphics[width=1\linewidth]{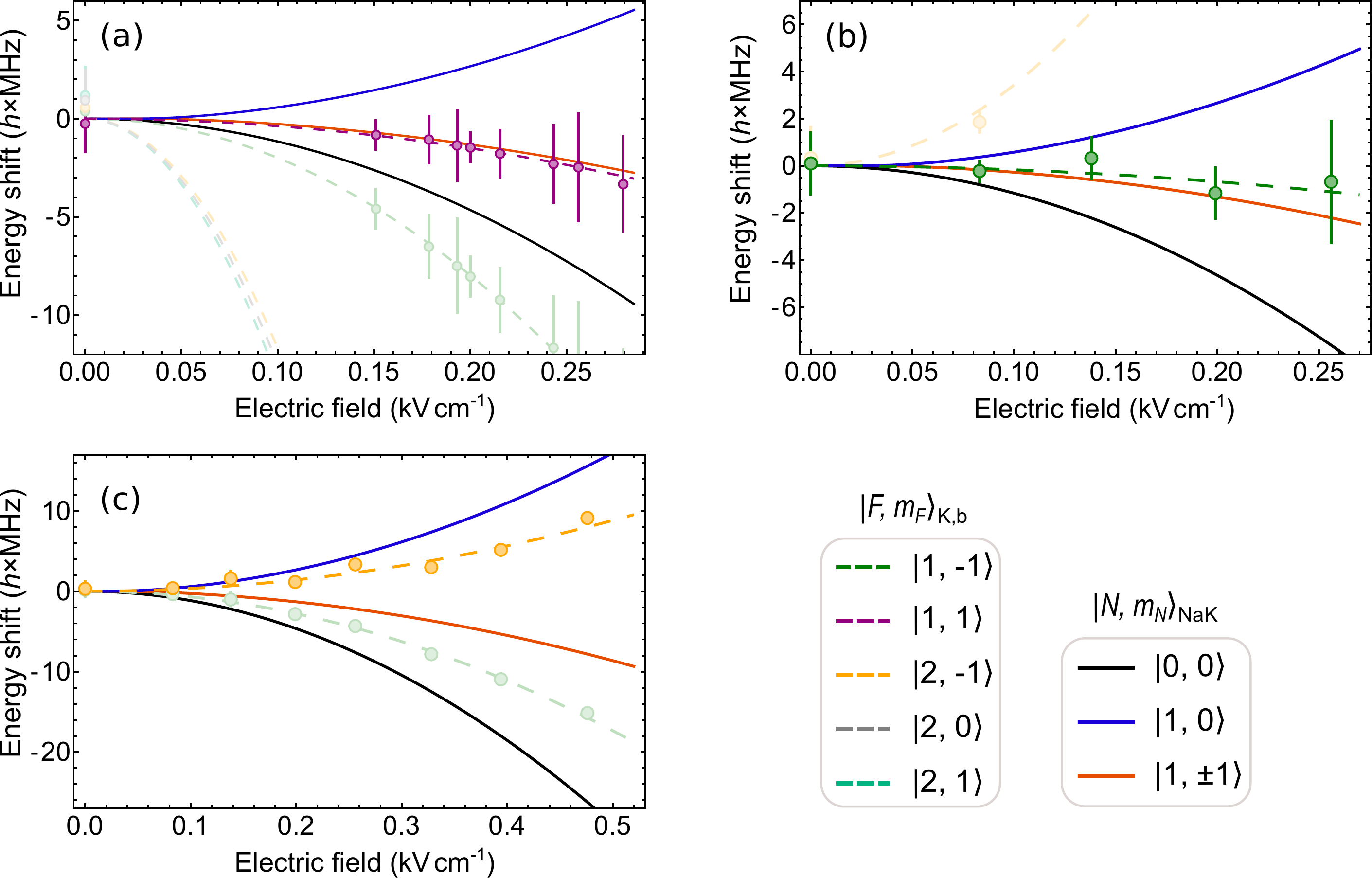}
\caption{
    Stark shifts of the bound state. Solid lines show the electric field dependence $U^{\textrm{NaK}}_{ N, m_N}(E)$ of the respective NaK states. 
    Points indicate the extracted energy shifts of the bound trimer state at $B=0$, assuming the K contribution as specified in the legend. Error bars represent the standard errors. Dashed lines are fits of the bound-state energy as a function of electric field using Eq. \eqref{eq:Fitformular} 
    for $N=0$, $m_N=0$. 
    Data and fits are offset by $U^{\textrm{b}}_{F,m_F}(0)$. All fit parameters are listed in Table \ref{tab:FitPar}. 
    Panels (a) and (b) correspond to the resonance measured in $\ket{1, 0}_{\textrm{K,s}}$, with (b) providing a magnified view. Panels (c) and (d) correspond to the two resonances in $\ket{2, -2}_{\textrm{K,s}}$: H and L, respectively (see Fig. \ref{fig:EfieldData} (b)).
}
\label{fig:EnergiesMain}
\end{figure*}

For an in-depth methodological discussion, please see Appendix \ref{app:DetailedDiscussion}. 
The energies of the bound states we have attributed to these resonances and the fits to Eq. \eqref{eq:Fitformular} are illustrated in Figure \ref{fig:EnergiesMain}.

\subsection*{Conclusion}
We have demonstrated electric-field control of Feshbach resonances in ultracold atom-molecule collisions. This establishes electric fields, alongside magnetic fields, as a new tool to manipulate interactions at the quantum level. By resolving the field dependence of the bound states, we were able to discuss the dimer rotation and the hyperfine component of K in the trimer state and to assign it unambiguously.\\
In one case (resonance L), we found a clear deviation of the Stark shifts from that of the uncoupled diatom. The bound-state response, in this instance, cannot be captured within the uncoupled-dimer picture, indicating additional constraints on the diatomic rotation inside the trimer.
Consequently, K can exert a significant influence on the diatom, even though the state is weakly bound. This behavior is consistent with hindered rotation: ab initio calculations indicate that the weakly bound NaK-K complex favors specific orientations, increasing rigidity and modifying the Stark response \cite{Shammout2025}.
In this case, bound states which were excluded in Appendix \ref{app:DetailedDiscussion} would also be possible.\\
These findings provide direct spectroscopic access to the internal makeup of triatomic states and highlight physical ingredients that must be incorporated into future theoretical descriptions of this kind of Feshbach resonances.\\
Beyond the specific \NaK–\K\ system, the approach we present can be generalized to other species, opening a versatile route to probing and controlling polyatomic resonances across a wide range of ultracold mixtures. This paves the way toward creating long-lived ensembles of trimers, refining models of polyatomic chemistry at ultralow temperatures, and ultimately engineering new forms of controlled polyatomic quantum matter.

\begin{acknowledgments}
We thank Philipp Gersema for his effort in improving the magnetic field stabilization, which made these measurements possible.
L.K. thanks the Deutsche Forschungsgemeinschaft (DFG, German Research Foundation) for support through the Heisenberg Programme No. 506287139.
M. M., J. H., F. G., B. S., L.K. and S.O. gratefully acknowledge financial support from the Deutsche Forschungsgemeinschaft (DFG, German Research Foundation) through CRC 1227 (DQ-mat), Project No. A03, and under Germany’s Excellence Strategy - EXC-2123 QuantumFrontiers - 390837967, and the European Research Council through ERC Consolidator Grant No. 101045075 - TRITRAMO.\\

M.M., J.H., F.G., L.K., and S.O. conceived the research; M.M., J.H., and F.G. performed the experiments and maintained the laboratory, with support from L.K.; The initial data analysis was conducted by M.M., J.H., and F.G.; M.M. performed the subsequent calculations of Stark shifts and, under E.T.’s supervision, conceptualized and executed the potential bound-state assignments; these assignments were discussed extensively and agreed upon by all authors; B.S. and E.T. provided theoretical insights into the trimer structure; L.K. and S.O. supervised and coordinated the research and resources; M.M. drafted the manuscript and prepared the visualizations. All authors contributed substantially to writing and revising the manuscript.
\end{acknowledgments}

\appendix

\section{\label{app:ExpDetails}Experimental details}

\begin{figure}[h]
    \centering
    \includegraphics[width=1\linewidth]{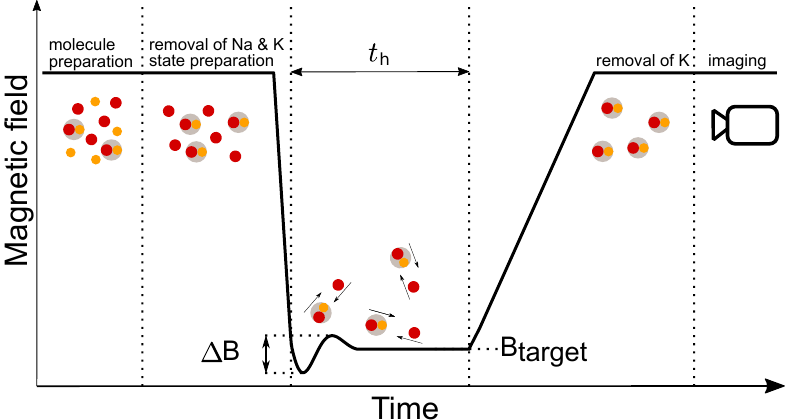}
    \caption{Experimental sequence. After molecule preparation, sodium is removed and potassium is prepared in a selected hyperfine state. The magnetic field is rapidly ramped to $B_{\textrm{target}}$, held for a time $t_{\textrm{h}}$, and then ramped back in a controlled manner to remove potassium once the field is stable and to image the remaining molecules. The fast ramp to $B_{\textrm{target}}$ leads to overshoot, which broadens the observed resonances by up to $\Delta B=\SI{0.4}{G}$.} 
    \label{fig:Sequence}
\end{figure}
We use the setup described in \cite{NaKmixture} to prepare a dense sample of ultracold sodium and potassium atoms in a crossed optical dipole trap. Ground-state molecules are then created via a controlled chemical reaction, as described in \cite{Kai}. \\
We obtain an ensemble of $\num{2.0 \pm 0.5 e5}$ potassium atoms in the hyperfine state $\ket{1,-1}_{\textrm{K}}$, $\num{5.0 \pm 0.5 e5}$ sodium atoms in the same hyperfine state and $\num{8 \pm 2 e3}{}$ NaK molecules in the rovibronic ground state and hyperfine state $\ket{m_{I,\mathrm{Na}}=-3/2, m_{I,\mathrm{K}}=-1/2}$ at a temperature of $T=\SI{520\pm 50}{nK}$ in a crossed optical dipole trap with trap frequencies $\omega_x=\SI{239\pm 10}{Hz},\ \omega_y=\SI{35\pm 5}{Hz},\ \omega_z=\SI{230\pm 10}{Hz}$ for NaK. 
As visualized in Fig. \ref{fig:Sequence}, the sodium atoms, which are still present in the mixture, are immediately removed after the creation of the ground-state molecules. The potassium atoms, which are initially in the state $\ket{F=1,\,m_F=-1}$, are  optionally transferred to another selected hyperfine state. We apply a fixed homogeneous electric field $E$ parallel to the magnetic field $B$ throughout the whole sequence. The setup for the electric fields is described in \cite{Gempel2016}. After preparing the mixture, $B$ is ramped to $B_{\textrm{target}}$, held for a time $t_{\textrm{h}}$ varying between $\SI{10}{ms}$ and $\SI{15}{ms}$ for the individual measurements 
and then ramped back to the initial value for removal of the K atoms and imaging of the molecules.
The target magnetic field has an uncertainty of $\Delta B=\SI{0.4}{G}$, due to overshooting during the sequence.
Since the electric field is applied throughout the entire measurement and calibrated for each measurement by means of a rotational transfer of the dimer, its relative accuracy is comparatively high and is in the order of $\SI{1}{V\,cm^{-1}} $ for the highest fields. This uncertainty is dominated by drifts due to moving charges in the experimental apparatus. However, the total uncertainty is limited by the uncertainty of the dipole moment of NaK, which is $2~\%$ \cite{Gerdes2011}. The fitted ratio of the squared dipole moment is not influenced by this.

\section{\label{app:DetailedDiscussion} Assignment of the bound states}

\subsection{Resonance in $\ket{F=1, m_F=0}_{\textrm{K,s}}$}

\begin{table*}
\caption{Fit parameters for bound state energies. Energies are referenced as in Fig. \ref{fig:DetermineEb} and fitted to Eq. \eqref{eq:Fitformular}. The asymptotic energies of the bound potassium for zero magnetic field $U^{\textrm{K}}_F/h$ are given for estimation of the binding energy according to the different cases of rotation of the molecular constituent. The indices H and L denote the resonances measured in higher and lower magnetic field for the channel $\ket{2, -2}_{\textrm{K,s}}$. The relative Stark shift $({\mu ^{\textrm{rel}}_{F,m_F, N, m_N}})^2$ referred to the molecular state is only displayed if positive, otherwise it is stated as not physical (n.p.). The stated uncertainties correspond to the standard error. The bound trimer states that we have ruled out are shown in gray.}
\label{tab:FitPar}%
\begin{ruledtabular}
\begin{tabular}{llccccc}
  $\ket{F,m_F}_{\textrm{K,s}}$&
     \textrm{$\ket{F,m_F}_{\textrm{K,b}}$}&
     \textrm{$U^{\textrm{K}}_F/h$ (MHz)}&
     \textrm{$U^{\textrm {b}}_{F,m_F}(0)/h$} \ ($\textrm{MHz}$)&
     \textrm{$({\mu ^{\textrm{rel}}_{F,m_F, 0,0}})^2$}&
      \textrm{$({\mu ^{\textrm{rel}}_{F,m_F, 1,\pm 1}})^2$}& \textrm{$({\mu ^{\textrm{rel}}_{F,m_F, 1, 0}})^2$}\\
      \hline
$\ket{1, 0}$ &  $\ket{1, 1}$ & $-288.57$ & $\num{-208.86 \pm 0.10}$ & $\num{0.32 \pm 0.02}$ & $\num{1.11 \pm 0.07}$ & \text{n.p.}\\
     \textcolor{gray}{$\ket{1, 0}$} & \textcolor{gray}{$\ket{1, -1}$} & \textcolor{gray}{$-288.57$} & \textcolor{gray}{$\num{-392.06 \pm 0.14}$ }& \textcolor{gray}{$\num{1.71 \pm 0.03}$}& \textcolor{gray}{$\num{5.8 \pm 0.2}$} & \textcolor{gray}{\text{n.p.}}\\
     \textcolor{gray}{$\ket{1, 0}$ } & \textcolor{gray}{$\ket{2, 1}$} & \textcolor{gray}{$+173.14$ }& \textcolor{gray}{$\num{-623.5 \pm 0.7}$ }& \textcolor{gray}{$\num{11.4 \pm 0.2}$}& \textcolor{gray}{$\num{38.9 \pm 0.9}$}& \textcolor{gray}{\text{n.p.}}\\
     \textcolor{gray}{$\ket{1, 0}$} & \textcolor{gray}{$\ket{2, 0}$} & \textcolor{gray}{$+173.14$ }&\textcolor{gray}{ $\num{-543.9 \pm 0.6}$ }& \textcolor{gray}{$\num{10.76 \pm 0.15}$}& \textcolor{gray}{$\num{36.7 \pm 0.8}$}& \textcolor{gray}{\text{n.p.}}\\
     \textcolor{gray}{$\ket{1, 0}$} & \textcolor{gray}{$\ket{2, -1}$} & \textcolor{gray}{$+173.14$} & \textcolor{gray}{$\num{-440.5 \pm 0.6}$} & \textcolor{gray}{$\num{10.05 \pm 0.13}$}& \textcolor{gray}{$\num{34.2 \pm 0.7}$ }& \textcolor{gray}{\text{n.p.}}
     \vspace{.06in}\\
     $\ket{2, -2}_{\textrm{H}}$ & $\ket{1, -1}$  & $-288.57$ & $\num{-82.7 \pm 0.3}$ & $\num{0.15 \pm 0.12}$ & $\num{0.5 \pm 0.4}$& \text{n.p.}\\
     \textcolor{gray}{$\ket{2, -2}_{\textrm{H}}$ }& \textcolor{gray}{$\ket{2, -1}$} & \textcolor{gray}{$+173.14$} & \textcolor{gray}{$\num{-153.2 \pm 1.2}$} & \textcolor{gray}{\text{n.p.}}& \textcolor{gray}{\text{n.p.}} & \textcolor{gray}{$\num{5.0 \pm 1.0}$}
     \vspace{.06in}\\
     \textcolor{gray}{$\ket{2, -2}_{\textrm{L}}$} & \textcolor{gray}{$\ket{1, -1}$} & \textcolor{gray}{$-288.57$} & \textcolor{gray}{$\num{-67.2 \pm 0.2}$} & \textcolor{gray}{$\num{0.599 \pm 0.014}$}& \textcolor{gray}{$\num{2.01 \pm 0.04}$}& \textcolor{gray}{\text{n.p.}}\\
     $\ket{2, -2}_{\textrm{L}}$ & $\ket{2, -1}$ & $+173.14$ & $\num{-88.4 \pm 0.4}$ & \text{n.p.} & \text{n.p.} &$\num{0.51\pm 0.05}$\\

\end{tabular}
\end{ruledtabular}
\end{table*}

First, we consider the scattering channel $\ket{1,0}_{\textrm{K,s}}$, which yields five possible bound states with $\ket{1,\pm 1}_{\textrm{K,b}},\ \ket{2,\pm 1}_{\textrm{K,b}}$ or $\ket{2,0}_{\textrm{K,b}}$, and we show the fit results to eq. \ref{eq:Fitformular} in Fig. \ref{fig:Energies}(a). The three states with $F=2$ would imply $({\mu ^{\textrm{rel}}_{2,m_F, 0, 0}})^2>10$, a Stark shift higher than ten times the one of the diatom. Under the assumption that the diatom rotates almost freely in the presence of K, this is unphysically large and such an assignment should be discarded. The two remaining candidates are enlarged in Fig. \ref{fig:Energies} (b). Assuming the bound state with $\ket{1, -1}_{\textrm{K,b}}$ implies $({\mu ^{\textrm{rel}}_{1,-1, 0, 0}})^2 = \num{1.71\pm 0.03}$ or $({\mu ^{\textrm{rel}}_{1,-1, 1,\pm 1}})^2 =\num{5.8 \pm 0.2}$ relative to the ground and first rotationally excited state, respectively. In this picture, these are too large to be explained by a polarization effect of NaK by a weakly bound K atom. In contrast, the remaining case $\ket{1, 1}_{\textrm{K,b}}$  yields $({\mu ^{\textrm{rel}}_{1,1, 1, \pm 1}})^2 = \num{1.11\pm 0.07}$, which is close  to the behavior of unperturbed NaK in the rotational state $\ket{1, \pm 1}_{\textrm{NaK}}$ ($1.6~\sigma$).
To fulfill angular momentum conservation, we assign the state $\ket{N=1,m_N=-1}_{\textrm{NaK}}\ket{F=1,m_F=1}_{\textrm{K}}$, without altering the hyperfine coupling in the diatomic part.\\

\subsection{Resonances in $\ket{F=2, m_F=-2}_{\textrm{K,s}}$}
\begin{figure*}[]
\includegraphics[width=1\textwidth]{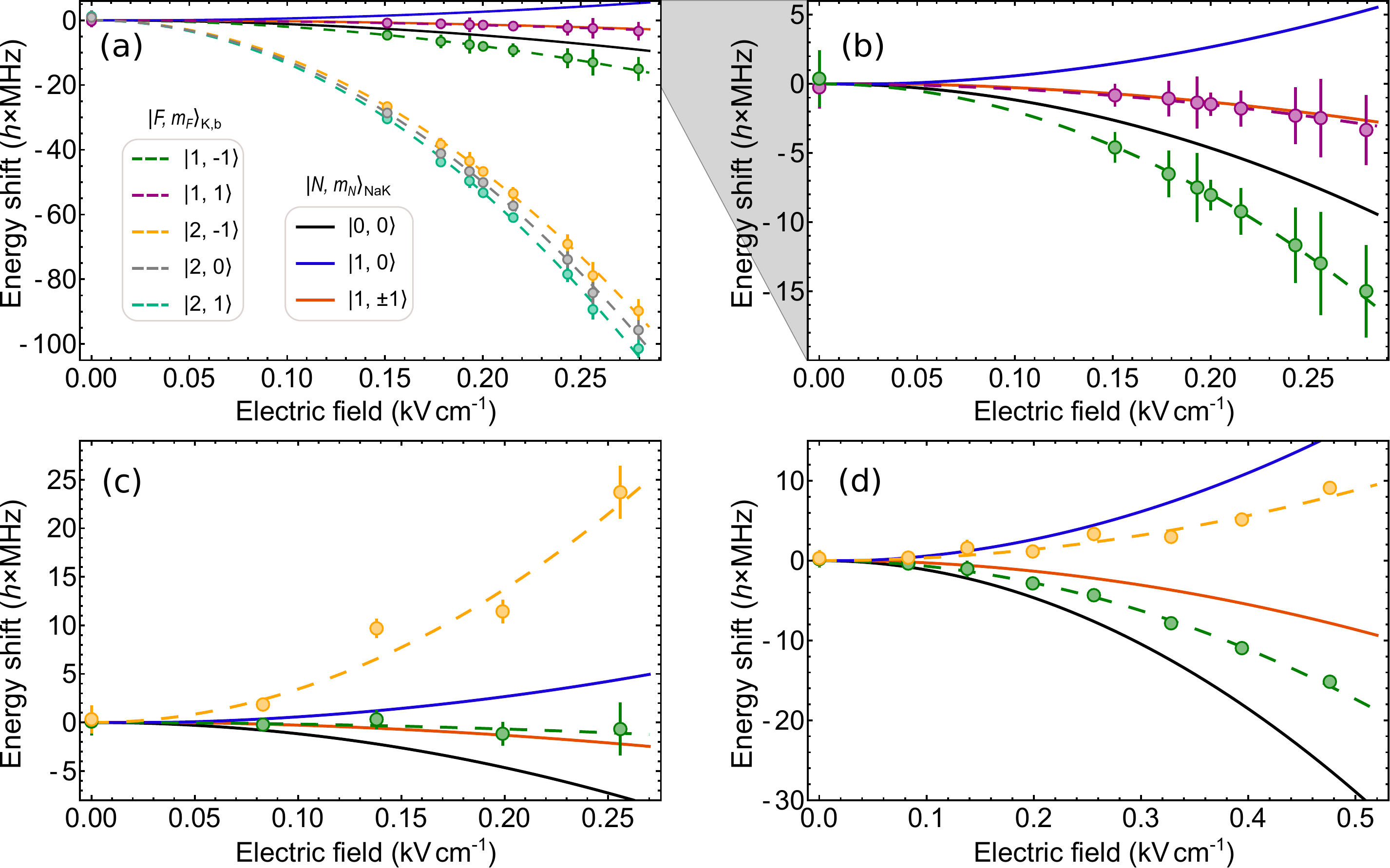}
\caption{Stark shifts of the bound state. Solid lines show the electric field dependence $U^{\textrm{NaK}}_{ N, m_N}(E)$ of the respective NaK states. 
    Points indicate the extracted energy shifts of the bound trimer state at $B=0$, assuming the K contribution as specified in the legend. Error bars represent the standard errors. Dashed lines are fits of the bound-state energy as a function of electric field using Eq. \eqref{eq:Fitformular} 
    for $N=0$, $m_N=0$. 
    Data and fits are offset by $U^{\textrm{b}}_{F,m_F}(0)$. All fit parameters are listed in Table \ref{tab:FitPar}. 
    Panels (a) and (b) correspond to the resonance measured in $\ket{1, 0}_{\textrm{K,s}}$, with (b) providing a magnified view. Panels (c) and (d) correspond to the two resonances in $\ket{2, -2}_{\textrm{K,s}}$: H and L, respectively (see Fig. \ref{fig:EfieldData} (b)).
}
\label{fig:Energies}
\end{figure*}

We repeat our assignment procedure for the two distinct Feshbach resonances emerging in the $\ket{2, -2}_{\textrm{K,s}}$ scattering state, namely the one appearing at higher magnetic field, designated as H (orange data points in Fig. \ref{fig:EfieldData} (b)) and L, appearing at lower magnetic field (blue data points in Fig. \ref{fig:EfieldData} (b)). We obtain two possible K bound states, which are $\ket{1, -1}_{\textrm{K,b}}$ and $\ket{2, -1}_{\textrm{K,b}}$, for each of the observed Feshbach resonances and calculate the corresponding Stark shifts depicted in Fig. \ref{fig:Energies} (c, d).\\

In the case of resonance H, we find that $\ket{2, -1}_{\textrm{K,b}}$ would imply $({\mu ^{\textrm{rel}}_{2,-1, 1, 0}})^2=\num{5.0\pm 1.0}$.
Because of this large magnitude, we exclude this option and identify the hyperfine state of K as $\ket{1, -1}_{\textrm{K,b}}$, for which we obtain $({\mu ^{\textrm{rel}}_{1,-1, 0, 0}})^2=\num{0.15 \pm 0.12}$ or $({\mu ^{\textrm{rel}}_{1,-1, 1, \pm 1}})^2=\num{0.5\pm 0.4}$.

Here the energy $U^{\mathrm{b}}_{F,m_F}(0)$ of the bound state is higher than the asymptotic energy $U_F^{\textrm{K}}$ of the bound potassium. Therefore, the asymptotic energy of the trimer needs to include the rotational energy of the dimer of $\approx h \times 6~\mathrm{GHz}$ to ensure that the state is bound, restricting the possible states to options  with $N=1$. To fulfill angular momentum conservation, we assign $\ket{N=1,m_N=-1}_{\mathrm{NaK}}\ket{F=1,m_F=-1}_{\mathrm{K}}$.\\


For resonance L, the first option $\ket{1, -1}_{\textrm{K,b}}$ is constrained to a combination with \mbox{$N=0$}, since \mbox{$N=1$} can be excluded due to its large corresponding relative Stark shift of $({\mu ^{\textrm{rel}}_{1,-1, 1, \pm 1}})^2=\num{2.01 \pm 0.04}$. The other option $\ket{2, -1}_{\textrm{K,b}}$ requires a rotation of the dimer since only the relative Stark shift for $\ket{N=1,m_N=0}$ is positive. 
Under this restriction, we compare the energies of the two remaining options with their respective asymptotes and find that only one of the states is actually bound.
Therefore, we assign a bound state with $\ket{N=1,m_N=0}_{\textrm{NaK}}\ket{F=2,m_F=-1}_{\textrm{K}}$ to the resonance L. As our measurement is not sensitive to the nuclear spin of the dimer or the overall rotation of the dimer and K, these remaining degrees of freedom allow the conservation of angular momentum of this assignment.

\bibliography{bibliography}

\begin{thebibliography}{31}%
\makeatletter
\providecommand \@ifxundefined [1]{%
 \@ifx{#1\undefined}
}%
\providecommand \@ifnum [1]{%
 \ifnum #1\expandafter \@firstoftwo
 \else \expandafter \@secondoftwo
 \fi
}%
\providecommand \@ifx [1]{%
 \ifx #1\expandafter \@firstoftwo
 \else \expandafter \@secondoftwo
 \fi
}%
\providecommand \natexlab [1]{#1}%
\providecommand \enquote  [1]{``#1''}%
\providecommand \bibnamefont  [1]{#1}%
\providecommand \bibfnamefont [1]{#1}%
\providecommand \citenamefont [1]{#1}%
\providecommand \href@noop [0]{\@secondoftwo}%
\providecommand \href [0]{\begingroup \@sanitize@url \@href}%
\providecommand \@href[1]{\@@startlink{#1}\@@href}%
\providecommand \@@href[1]{\endgroup#1\@@endlink}%
\providecommand \@sanitize@url [0]{\catcode `\\12\catcode `\$12\catcode
  `\&12\catcode `\#12\catcode `\^12\catcode `\_12\catcode `\%12\relax}%
\providecommand \@@startlink[1]{}%
\providecommand \@@endlink[0]{}%
\providecommand \url  [0]{\begingroup\@sanitize@url \@url }%
\providecommand \@url [1]{\endgroup\@href {#1}{\urlprefix }}%
\providecommand \urlprefix  [0]{URL }%
\providecommand \Eprint [0]{\href }%
\providecommand \doibase [0]{https://doi.org/}%
\providecommand \selectlanguage [0]{\@gobble}%
\providecommand \bibinfo  [0]{\@secondoftwo}%
\providecommand \bibfield  [0]{\@secondoftwo}%
\providecommand \translation [1]{[#1]}%
\providecommand \BibitemOpen [0]{}%
\providecommand \bibitemStop [0]{}%
\providecommand \bibitemNoStop [0]{.\EOS\space}%
\providecommand \EOS [0]{\spacefactor3000\relax}%
\providecommand \BibitemShut  [1]{\csname bibitem#1\endcsname}%
\let\auto@bib@innerbib\@empty
\bibitem [{\citenamefont {Cornish}\ \emph {et~al.}(2024)\citenamefont
  {Cornish}, \citenamefont {Tarbutt},\ and\ \citenamefont
  {Hazzard}}]{Cornish2024}%
  \BibitemOpen
  \bibfield  {author} {\bibinfo {author} {\bibfnamefont {S.~L.}\ \bibnamefont
  {Cornish}}, \bibinfo {author} {\bibfnamefont {M.~R.}\ \bibnamefont
  {Tarbutt}},\ and\ \bibinfo {author} {\bibfnamefont {K.~R.~A.}\ \bibnamefont
  {Hazzard}},\ }\bibfield  {title} {\bibinfo {title} {{Quantum computation and
  quantum simulation with ultracold molecules}},\ }\href
  {https://doi.org/10.1038/s41567-024-02453-9} {\bibfield  {journal} {\bibinfo
  {journal} {\textit{Nat. Phys.}}\ }\textbf {\bibinfo {volume} {20}},\ \bibinfo
  {pages} {730–740} (\bibinfo {year} {2024})}\BibitemShut {NoStop}%
\bibitem [{\citenamefont {Bohn}\ \emph {et~al.}(2017)\citenamefont {Bohn},
  \citenamefont {Rey},\ and\ \citenamefont {Ye}}]{Bohn2017}%
  \BibitemOpen
  \bibfield  {author} {\bibinfo {author} {\bibfnamefont {J.~L.}\ \bibnamefont
  {Bohn}}, \bibinfo {author} {\bibfnamefont {A.~M.}\ \bibnamefont {Rey}},\ and\
  \bibinfo {author} {\bibfnamefont {J.}~\bibnamefont {Ye}},\ }\bibfield
  {title} {\bibinfo {title} {{Cold molecules: Progress in quantum engineering
  of chemistry and quantum matter}},\ }\href
  {https://doi.org/10.1126/science.aam6299} {\bibfield  {journal} {\bibinfo
  {journal} {\textit{Science}}\ }\textbf {\bibinfo {volume} {357}},\ \bibinfo
  {pages} {1002–1010} (\bibinfo {year} {2017})}\BibitemShut {NoStop}%
\bibitem [{\citenamefont {Karman}\ and\ \citenamefont
  {Hutson}(2018)}]{MWshieldProposalKarman}%
  \BibitemOpen
  \bibfield  {author} {\bibinfo {author} {\bibfnamefont {T.}~\bibnamefont
  {Karman}}\ and\ \bibinfo {author} {\bibfnamefont {J.~M.}\ \bibnamefont
  {Hutson}},\ }\bibfield  {title} {\bibinfo {title} {{Microwave Shielding of
  Ultracold Polar Molecules}},\ }\href
  {https://doi.org/10.1103/PhysRevLett.121.163401} {\bibfield  {journal}
  {\bibinfo  {journal} {\textit{Phys. Rev. Lett.}}\ }\textbf {\bibinfo {volume}
  {121}},\ \bibinfo {pages} {163401} (\bibinfo {year} {2018})}\BibitemShut
  {NoStop}%
\bibitem [{\citenamefont {Lassablière}\ and\ \citenamefont
  {Quéméner}(2018)}]{MWshieldProposalQuemener}%
  \BibitemOpen
  \bibfield  {author} {\bibinfo {author} {\bibfnamefont {L.}~\bibnamefont
  {Lassablière}}\ and\ \bibinfo {author} {\bibfnamefont {G.}~\bibnamefont
  {Quéméner}},\ }\bibfield  {title} {\bibinfo {title} {{Controlling the
  Scattering Length of Ultracold Dipolar Molecules}},\ }\href
  {https://doi.org/10.1103/physrevlett.121.163402} {\bibfield  {journal}
  {\bibinfo  {journal} {\textit{Phys. Rev. Lett.}}\ }\textbf {\bibinfo {volume}
  {121}},\ \bibinfo {pages} {163402} (\bibinfo {year} {2018})}\BibitemShut
  {NoStop}%
\bibitem [{\citenamefont {Matsuda}\ \emph {et~al.}(2020)\citenamefont
  {Matsuda}, \citenamefont {De~Marco}, \citenamefont {Li}, \citenamefont
  {Tobias}, \citenamefont {Valtolina}, \citenamefont {Quéméner},\ and\
  \citenamefont {Ye}}]{EfieldShielding}%
  \BibitemOpen
  \bibfield  {author} {\bibinfo {author} {\bibfnamefont {K.}~\bibnamefont
  {Matsuda}}, \bibinfo {author} {\bibfnamefont {L.}~\bibnamefont {De~Marco}},
  \bibinfo {author} {\bibfnamefont {J.-R.}\ \bibnamefont {Li}}, \bibinfo
  {author} {\bibfnamefont {W.~G.}\ \bibnamefont {Tobias}}, \bibinfo {author}
  {\bibfnamefont {G.}~\bibnamefont {Valtolina}}, \bibinfo {author}
  {\bibfnamefont {G.}~\bibnamefont {Quéméner}},\ and\ \bibinfo {author}
  {\bibfnamefont {J.}~\bibnamefont {Ye}},\ }\bibfield  {title} {\bibinfo
  {title} {{Resonant collisional shielding of reactive molecules using electric
  fields}},\ }\href {https://doi.org/10.1126/science.abe7370} {\bibfield
  {journal} {\bibinfo  {journal} {\textit{Science}}\ }\textbf {\bibinfo
  {volume} {370}},\ \bibinfo {pages} {1324–1327} (\bibinfo {year}
  {2020})}\BibitemShut {NoStop}%
\bibitem [{\citenamefont {Karam}\ \emph {et~al.}(2023)\citenamefont {Karam},
  \citenamefont {Vexiau}, \citenamefont {Bouloufa-Maafa}, \citenamefont
  {Dulieu}, \citenamefont {Lepers}, \citenamefont {Meyer~zum Alten~Borgloh},
  \citenamefont {Ospelkaus},\ and\ \citenamefont {Karpa}}]{Karam2023}%
  \BibitemOpen
  \bibfield  {author} {\bibinfo {author} {\bibfnamefont {C.}~\bibnamefont
  {Karam}}, \bibinfo {author} {\bibfnamefont {R.}~\bibnamefont {Vexiau}},
  \bibinfo {author} {\bibfnamefont {N.}~\bibnamefont {Bouloufa-Maafa}},
  \bibinfo {author} {\bibfnamefont {O.}~\bibnamefont {Dulieu}}, \bibinfo
  {author} {\bibfnamefont {M.}~\bibnamefont {Lepers}}, \bibinfo {author}
  {\bibfnamefont {M.}~\bibnamefont {Meyer~zum Alten~Borgloh}}, \bibinfo
  {author} {\bibfnamefont {S.}~\bibnamefont {Ospelkaus}},\ and\ \bibinfo
  {author} {\bibfnamefont {L.}~\bibnamefont {Karpa}},\ }\bibfield  {title}
  {\bibinfo {title} {{Two-photon optical shielding of collisions between
  ultracold polar molecules}},\ }\href
  {https://doi.org/10.1103/physrevresearch.5.033074} {\bibfield  {journal}
  {\bibinfo  {journal} {\textit{Phys. Rev. Res.}}\ }\textbf {\bibinfo {volume}
  {5}},\ \bibinfo {pages} {033074} (\bibinfo {year} {2023})}\BibitemShut
  {NoStop}%
\bibitem [{\citenamefont {Anderegg}\ \emph {et~al.}(2021)\citenamefont
  {Anderegg}, \citenamefont {Burchesky}, \citenamefont {Bao}, \citenamefont
  {Yu}, \citenamefont {Karman}, \citenamefont {Chae}, \citenamefont {Ni},
  \citenamefont {Ketterle},\ and\ \citenamefont {Doyle}}]{Anderegg2021}%
  \BibitemOpen
  \bibfield  {author} {\bibinfo {author} {\bibfnamefont {L.}~\bibnamefont
  {Anderegg}}, \bibinfo {author} {\bibfnamefont {S.}~\bibnamefont {Burchesky}},
  \bibinfo {author} {\bibfnamefont {Y.}~\bibnamefont {Bao}}, \bibinfo {author}
  {\bibfnamefont {S.~S.}\ \bibnamefont {Yu}}, \bibinfo {author} {\bibfnamefont
  {T.}~\bibnamefont {Karman}}, \bibinfo {author} {\bibfnamefont
  {E.}~\bibnamefont {Chae}}, \bibinfo {author} {\bibfnamefont {K.-K.}\
  \bibnamefont {Ni}}, \bibinfo {author} {\bibfnamefont {W.}~\bibnamefont
  {Ketterle}},\ and\ \bibinfo {author} {\bibfnamefont {J.~M.}\ \bibnamefont
  {Doyle}},\ }\bibfield  {title} {\bibinfo {title} {{Observation of microwave
  shielding of ultracold molecules}},\ }\href
  {https://doi.org/10.1126/science.abg9502} {\bibfield  {journal} {\bibinfo
  {journal} {\textit{Science}}\ }\textbf {\bibinfo {volume} {373}},\ \bibinfo
  {pages} {779–782} (\bibinfo {year} {2021})}\BibitemShut {NoStop}%
\bibitem [{\citenamefont {Bigagli}\ \emph {et~al.}(2023)\citenamefont
  {Bigagli}, \citenamefont {Warner}, \citenamefont {Yuan}, \citenamefont
  {Zhang}, \citenamefont {Stevenson}, \citenamefont {Karman},\ and\
  \citenamefont {Will}}]{ShieldingWill}%
  \BibitemOpen
  \bibfield  {author} {\bibinfo {author} {\bibfnamefont {N.}~\bibnamefont
  {Bigagli}}, \bibinfo {author} {\bibfnamefont {C.}~\bibnamefont {Warner}},
  \bibinfo {author} {\bibfnamefont {W.}~\bibnamefont {Yuan}}, \bibinfo {author}
  {\bibfnamefont {S.}~\bibnamefont {Zhang}}, \bibinfo {author} {\bibfnamefont
  {I.}~\bibnamefont {Stevenson}}, \bibinfo {author} {\bibfnamefont
  {T.}~\bibnamefont {Karman}},\ and\ \bibinfo {author} {\bibfnamefont
  {S.}~\bibnamefont {Will}},\ }\bibfield  {title} {\bibinfo {title}
  {{Collisionally stable gas of bosonic dipolar ground-state molecules}},\
  }\href {https://doi.org/10.1038/s41567-023-02200-6} {\bibfield  {journal}
  {\bibinfo  {journal} {\textit{Nat. Phys.}}\ }\textbf {\bibinfo {volume}
  {19}},\ \bibinfo {pages} {1579–1584} (\bibinfo {year} {2023})}\BibitemShut
  {NoStop}%
\bibitem [{\citenamefont {Lin}\ \emph {et~al.}(2023)\citenamefont {Lin},
  \citenamefont {Chen}, \citenamefont {Jin}, \citenamefont {Shi}, \citenamefont
  {Deng}, \citenamefont {Zhang}, \citenamefont {Quéméner}, \citenamefont
  {Shi}, \citenamefont {Yi},\ and\ \citenamefont {Wang}}]{ShieldingWang}%
  \BibitemOpen
  \bibfield  {author} {\bibinfo {author} {\bibfnamefont {J.}~\bibnamefont
  {Lin}}, \bibinfo {author} {\bibfnamefont {G.}~\bibnamefont {Chen}}, \bibinfo
  {author} {\bibfnamefont {M.}~\bibnamefont {Jin}}, \bibinfo {author}
  {\bibfnamefont {Z.}~\bibnamefont {Shi}}, \bibinfo {author} {\bibfnamefont
  {F.}~\bibnamefont {Deng}}, \bibinfo {author} {\bibfnamefont {W.}~\bibnamefont
  {Zhang}}, \bibinfo {author} {\bibfnamefont {G.}~\bibnamefont {Quéméner}},
  \bibinfo {author} {\bibfnamefont {T.}~\bibnamefont {Shi}}, \bibinfo {author}
  {\bibfnamefont {S.}~\bibnamefont {Yi}},\ and\ \bibinfo {author}
  {\bibfnamefont {D.}~\bibnamefont {Wang}},\ }\bibfield  {title} {\bibinfo
  {title} {{Microwave Shielding of Bosonic NaRb Molecules}},\ }\href
  {https://doi.org/10.1103/physrevx.13.031032} {\bibfield  {journal} {\bibinfo
  {journal} {\textit{Phys. Rev. X}}\ }\textbf {\bibinfo {volume} {13}},\
  \bibinfo {pages} {031032} (\bibinfo {year} {2023})}\BibitemShut {NoStop}%
\bibitem [{\citenamefont {Schindewolf}\ \emph {et~al.}(2022)\citenamefont
  {Schindewolf}, \citenamefont {Bause}, \citenamefont {Chen}, \citenamefont
  {Duda}, \citenamefont {Karman}, \citenamefont {Bloch},\ and\ \citenamefont
  {Luo}}]{FermigasMunich}%
  \BibitemOpen
  \bibfield  {author} {\bibinfo {author} {\bibfnamefont {A.}~\bibnamefont
  {Schindewolf}}, \bibinfo {author} {\bibfnamefont {R.}~\bibnamefont {Bause}},
  \bibinfo {author} {\bibfnamefont {X.-Y.}\ \bibnamefont {Chen}}, \bibinfo
  {author} {\bibfnamefont {M.}~\bibnamefont {Duda}}, \bibinfo {author}
  {\bibfnamefont {T.}~\bibnamefont {Karman}}, \bibinfo {author} {\bibfnamefont
  {I.}~\bibnamefont {Bloch}},\ and\ \bibinfo {author} {\bibfnamefont {X.-Y.}\
  \bibnamefont {Luo}},\ }\bibfield  {title} {\bibinfo {title} {{Evaporation of
  microwave-shielded polar molecules to quantum degeneracy}},\ }\href
  {https://doi.org/10.1038/s41586-022-04900-0} {\bibfield  {journal} {\bibinfo
  {journal} {\textit{Nature}}\ }\textbf {\bibinfo {volume} {607}},\ \bibinfo
  {pages} {677–681} (\bibinfo {year} {2022})}\BibitemShut {NoStop}%
\bibitem [{\citenamefont {Bigagli}\ \emph {et~al.}(2024)\citenamefont
  {Bigagli}, \citenamefont {Yuan}, \citenamefont {Zhang}, \citenamefont
  {Bulatovic}, \citenamefont {Karman}, \citenamefont {Stevenson},\ and\
  \citenamefont {Will}}]{BECwill}%
  \BibitemOpen
  \bibfield  {author} {\bibinfo {author} {\bibfnamefont {N.}~\bibnamefont
  {Bigagli}}, \bibinfo {author} {\bibfnamefont {W.}~\bibnamefont {Yuan}},
  \bibinfo {author} {\bibfnamefont {S.}~\bibnamefont {Zhang}}, \bibinfo
  {author} {\bibfnamefont {B.}~\bibnamefont {Bulatovic}}, \bibinfo {author}
  {\bibfnamefont {T.}~\bibnamefont {Karman}}, \bibinfo {author} {\bibfnamefont
  {I.}~\bibnamefont {Stevenson}},\ and\ \bibinfo {author} {\bibfnamefont
  {S.}~\bibnamefont {Will}},\ }\bibfield  {title} {\bibinfo {title}
  {{Observation of Bose–Einstein condensation of dipolar molecules}},\ }\href
  {https://doi.org/10.1038/s41586-024-07492-z} {\bibfield  {journal} {\bibinfo
  {journal} {\textit{Nature}}\ }\textbf {\bibinfo {volume} {631}},\ \bibinfo
  {pages} {289–293} (\bibinfo {year} {2024})}\BibitemShut {NoStop}%
\bibitem [{\citenamefont {Shi}\ \emph {et~al.}(2025)\citenamefont {Shi},
  \citenamefont {Huang}, \citenamefont {Deng}, \citenamefont {Jin},
  \citenamefont {Yi}, \citenamefont {Shi},\ and\ \citenamefont
  {Wang}}]{BECWangArxiv}%
  \BibitemOpen
  \bibfield  {author} {\bibinfo {author} {\bibfnamefont {Z.}~\bibnamefont
  {Shi}}, \bibinfo {author} {\bibfnamefont {Z.}~\bibnamefont {Huang}}, \bibinfo
  {author} {\bibfnamefont {F.}~\bibnamefont {Deng}}, \bibinfo {author}
  {\bibfnamefont {W.-J.}\ \bibnamefont {Jin}}, \bibinfo {author} {\bibfnamefont
  {S.}~\bibnamefont {Yi}}, \bibinfo {author} {\bibfnamefont {T.}~\bibnamefont
  {Shi}},\ and\ \bibinfo {author} {\bibfnamefont {D.}~\bibnamefont {Wang}},\
  }\bibfield  {title} {\bibinfo {title} {{Bose-Einstein condensate of ultracold
  sodium-rubidium molecules with tunable dipolar interactions}},\ }\href
  {https://doi.org/10.48550/ARXIV.2508.20518} {\bibfield  {journal} {\bibinfo
  {journal} {\textit{arXiv}}\ }\textbf {\bibinfo {volume} {2508.20518}}
  (\bibinfo {year} {2025})}\BibitemShut {NoStop}%
\bibitem [{\citenamefont {Yang}\ \emph {et~al.}(2022)\citenamefont {Yang},
  \citenamefont {Cao}, \citenamefont {Su}, \citenamefont {Rui}, \citenamefont
  {Zhao},\ and\ \citenamefont {Pan}}]{TrimersPan}%
  \BibitemOpen
  \bibfield  {author} {\bibinfo {author} {\bibfnamefont {H.}~\bibnamefont
  {Yang}}, \bibinfo {author} {\bibfnamefont {J.}~\bibnamefont {Cao}}, \bibinfo
  {author} {\bibfnamefont {Z.}~\bibnamefont {Su}}, \bibinfo {author}
  {\bibfnamefont {J.}~\bibnamefont {Rui}}, \bibinfo {author} {\bibfnamefont
  {B.}~\bibnamefont {Zhao}},\ and\ \bibinfo {author} {\bibfnamefont {J.-W.}\
  \bibnamefont {Pan}},\ }\bibfield  {title} {\bibinfo {title} {{Creation of an
  ultracold gas of triatomic molecules from an atom–diatomic molecule
  mixture}},\ }\href {https://doi.org/10.1126/science.ade6307} {\bibfield
  {journal} {\bibinfo  {journal} {\textit{Science}}\ }\textbf {\bibinfo
  {volume} {378}},\ \bibinfo {pages} {1009–1013} (\bibinfo {year}
  {2022})}\BibitemShut {NoStop}%
\bibitem [{\citenamefont {Chen}\ \emph {et~al.}(2024)\citenamefont {Chen},
  \citenamefont {Biswas}, \citenamefont {Eppelt}, \citenamefont {Schindewolf},
  \citenamefont {Deng}, \citenamefont {Shi}, \citenamefont {Yi}, \citenamefont
  {Hilker}, \citenamefont {Bloch},\ and\ \citenamefont
  {Luo}}]{FieldLinkedTetramerLuo}%
  \BibitemOpen
  \bibfield  {author} {\bibinfo {author} {\bibfnamefont {X.-Y.}\ \bibnamefont
  {Chen}}, \bibinfo {author} {\bibfnamefont {S.}~\bibnamefont {Biswas}},
  \bibinfo {author} {\bibfnamefont {S.}~\bibnamefont {Eppelt}}, \bibinfo
  {author} {\bibfnamefont {A.}~\bibnamefont {Schindewolf}}, \bibinfo {author}
  {\bibfnamefont {F.}~\bibnamefont {Deng}}, \bibinfo {author} {\bibfnamefont
  {T.}~\bibnamefont {Shi}}, \bibinfo {author} {\bibfnamefont {S.}~\bibnamefont
  {Yi}}, \bibinfo {author} {\bibfnamefont {T.~A.}\ \bibnamefont {Hilker}},
  \bibinfo {author} {\bibfnamefont {I.}~\bibnamefont {Bloch}},\ and\ \bibinfo
  {author} {\bibfnamefont {X.-Y.}\ \bibnamefont {Luo}},\ }\bibfield  {title}
  {\bibinfo {title} {{Ultracold field-linked tetratomic molecules}},\ }\href
  {https://doi.org/10.1038/s41586-023-06986-6} {\bibfield  {journal} {\bibinfo
  {journal} {\textit{Nature}}\ }\textbf {\bibinfo {volume} {626}},\ \bibinfo
  {pages} {283–287} (\bibinfo {year} {2024})}\BibitemShut {NoStop}%
\bibitem [{\citenamefont {Chin}\ \emph {et~al.}(2010)\citenamefont {Chin},
  \citenamefont {Grimm}, \citenamefont {Julienne},\ and\ \citenamefont
  {Tiesinga}}]{FeshMole}%
  \BibitemOpen
  \bibfield  {author} {\bibinfo {author} {\bibfnamefont {C.}~\bibnamefont
  {Chin}}, \bibinfo {author} {\bibfnamefont {R.}~\bibnamefont {Grimm}},
  \bibinfo {author} {\bibfnamefont {P.}~\bibnamefont {Julienne}},\ and\
  \bibinfo {author} {\bibfnamefont {E.}~\bibnamefont {Tiesinga}},\ }\bibfield
  {title} {\bibinfo {title} {{Feshbach resonances in ultracold gases}},\ }\href
  {https://doi.org/10.1103/revmodphys.82.1225} {\bibfield  {journal} {\bibinfo
  {journal} {\textit{Rev. Mod. Phys}}\ }\textbf {\bibinfo {volume} {82}},\
  \bibinfo {pages} {1225–1286} (\bibinfo {year} {2010})}\BibitemShut
  {NoStop}%
\bibitem [{\citenamefont {Son}\ \emph {et~al.}(2022)\citenamefont {Son},
  \citenamefont {Park}, \citenamefont {Lu}, \citenamefont {Jamison},
  \citenamefont {Karman},\ and\ \citenamefont
  {Ketterle}}]{ControllingCollisionsKetterle}%
  \BibitemOpen
  \bibfield  {author} {\bibinfo {author} {\bibfnamefont {H.}~\bibnamefont
  {Son}}, \bibinfo {author} {\bibfnamefont {J.~J.}\ \bibnamefont {Park}},
  \bibinfo {author} {\bibfnamefont {Y.-K.}\ \bibnamefont {Lu}}, \bibinfo
  {author} {\bibfnamefont {A.~O.}\ \bibnamefont {Jamison}}, \bibinfo {author}
  {\bibfnamefont {T.}~\bibnamefont {Karman}},\ and\ \bibinfo {author}
  {\bibfnamefont {W.}~\bibnamefont {Ketterle}},\ }\bibfield  {title} {\bibinfo
  {title} {{Control of reactive collisions by quantum interference}},\ }\href
  {https://doi.org/10.1126/science.abl7257} {\bibfield  {journal} {\bibinfo
  {journal} {\textit{Science}}\ }\textbf {\bibinfo {volume} {375}},\ \bibinfo
  {pages} {1006–1010} (\bibinfo {year} {2022})}\BibitemShut {NoStop}%
\bibitem [{\citenamefont {Karman}\ \emph {et~al.}(2023)\citenamefont {Karman},
  \citenamefont {Gronowski}, \citenamefont {Tomza}, \citenamefont {Park},
  \citenamefont {Son}, \citenamefont {Lu}, \citenamefont {Jamison},\ and\
  \citenamefont {Ketterle}}]{AbInitioKarman}%
  \BibitemOpen
  \bibfield  {author} {\bibinfo {author} {\bibfnamefont {T.}~\bibnamefont
  {Karman}}, \bibinfo {author} {\bibfnamefont {M.}~\bibnamefont {Gronowski}},
  \bibinfo {author} {\bibfnamefont {M.}~\bibnamefont {Tomza}}, \bibinfo
  {author} {\bibfnamefont {J.~J.}\ \bibnamefont {Park}}, \bibinfo {author}
  {\bibfnamefont {H.}~\bibnamefont {Son}}, \bibinfo {author} {\bibfnamefont
  {Y.-K.}\ \bibnamefont {Lu}}, \bibinfo {author} {\bibfnamefont {A.~O.}\
  \bibnamefont {Jamison}},\ and\ \bibinfo {author} {\bibfnamefont
  {W.}~\bibnamefont {Ketterle}},\ }\bibfield  {title} {\bibinfo {title} {{Ab
  initio calculation of the spectrum of Feshbach resonances in NaLi + Na
  collisions}},\ }\href {https://doi.org/10.1103/physreva.108.023309}
  {\bibfield  {journal} {\bibinfo  {journal} {\textit{Phys. Rev. A}}\ }\textbf
  {\bibinfo {volume} {108}},\ \bibinfo {pages} {023309} (\bibinfo {year}
  {2023})}\BibitemShut {NoStop}%
\bibitem [{\citenamefont {Park}\ \emph {et~al.}(2023)\citenamefont {Park},
  \citenamefont {Son}, \citenamefont {Lu}, \citenamefont {Karman},
  \citenamefont {Gronowski}, \citenamefont {Tomza}, \citenamefont {Jamison},\
  and\ \citenamefont {Ketterle}}]{TrimerFeshisKetterle}%
  \BibitemOpen
  \bibfield  {author} {\bibinfo {author} {\bibfnamefont {J.~J.}\ \bibnamefont
  {Park}}, \bibinfo {author} {\bibfnamefont {H.}~\bibnamefont {Son}}, \bibinfo
  {author} {\bibfnamefont {Y.-K.}\ \bibnamefont {Lu}}, \bibinfo {author}
  {\bibfnamefont {T.}~\bibnamefont {Karman}}, \bibinfo {author} {\bibfnamefont
  {M.}~\bibnamefont {Gronowski}}, \bibinfo {author} {\bibfnamefont
  {M.}~\bibnamefont {Tomza}}, \bibinfo {author} {\bibfnamefont {A.~O.}\
  \bibnamefont {Jamison}},\ and\ \bibinfo {author} {\bibfnamefont
  {W.}~\bibnamefont {Ketterle}},\ }\bibfield  {title} {\bibinfo {title}
  {{Spectrum of Feshbach Resonances in NaLi+Na Collisions}},\ }\href
  {https://doi.org/10.1103/physrevx.13.031018} {\bibfield  {journal} {\bibinfo
  {journal} {\textit{Phy. Rev. X}}\ }\textbf {\bibinfo {volume} {13}},\
  \bibinfo {pages} {031018} (\bibinfo {year} {2023})}\BibitemShut {NoStop}%
\bibitem [{\citenamefont {Yang}\ \emph {et~al.}(2019)\citenamefont {Yang},
  \citenamefont {Zhang}, \citenamefont {Liu}, \citenamefont {Liu},
  \citenamefont {Nan}, \citenamefont {Zhao},\ and\ \citenamefont
  {Pan}}]{Yang2019}%
  \BibitemOpen
  \bibfield  {author} {\bibinfo {author} {\bibfnamefont {H.}~\bibnamefont
  {Yang}}, \bibinfo {author} {\bibfnamefont {D.-C.}\ \bibnamefont {Zhang}},
  \bibinfo {author} {\bibfnamefont {L.}~\bibnamefont {Liu}}, \bibinfo {author}
  {\bibfnamefont {Y.-X.}\ \bibnamefont {Liu}}, \bibinfo {author} {\bibfnamefont
  {J.}~\bibnamefont {Nan}}, \bibinfo {author} {\bibfnamefont {B.}~\bibnamefont
  {Zhao}},\ and\ \bibinfo {author} {\bibfnamefont {J.-W.}\ \bibnamefont
  {Pan}},\ }\bibfield  {title} {\bibinfo {title} {{Observation of magnetically
  tunable Fesh\-bach resonances in ultracold $^{23}\textrm{Na}^{40}\textrm{K}$
  + $^{40}\textrm{K}$ collisions}},\ }\href
  {https://doi.org/10.1126/science.aau5322} {\bibfield  {journal} {\bibinfo
  {journal} {\textit{Science}}\ }\textbf {\bibinfo {volume} {363}},\ \bibinfo
  {pages} {261–264} (\bibinfo {year} {2019})}\BibitemShut {NoStop}%
\bibitem [{\citenamefont {Wang}\ \emph {et~al.}(2021)\citenamefont {Wang},
  \citenamefont {Frye}, \citenamefont {Su}, \citenamefont {Cao}, \citenamefont
  {Liu}, \citenamefont {Zhang}, \citenamefont {Yang}, \citenamefont {Hutson},
  \citenamefont {Zhao}, \citenamefont {Bai},\ and\ \citenamefont
  {Pan}}]{Wang2021}%
  \BibitemOpen
  \bibfield  {author} {\bibinfo {author} {\bibfnamefont {X.-Y.}\ \bibnamefont
  {Wang}}, \bibinfo {author} {\bibfnamefont {M.~D.}\ \bibnamefont {Frye}},
  \bibinfo {author} {\bibfnamefont {Z.}~\bibnamefont {Su}}, \bibinfo {author}
  {\bibfnamefont {J.}~\bibnamefont {Cao}}, \bibinfo {author} {\bibfnamefont
  {L.}~\bibnamefont {Liu}}, \bibinfo {author} {\bibfnamefont {D.-C.}\
  \bibnamefont {Zhang}}, \bibinfo {author} {\bibfnamefont {H.}~\bibnamefont
  {Yang}}, \bibinfo {author} {\bibfnamefont {J.~M.}\ \bibnamefont {Hutson}},
  \bibinfo {author} {\bibfnamefont {B.}~\bibnamefont {Zhao}}, \bibinfo {author}
  {\bibfnamefont {C.-L.}\ \bibnamefont {Bai}},\ and\ \bibinfo {author}
  {\bibfnamefont {J.-W.}\ \bibnamefont {Pan}},\ }\bibfield  {title} {\bibinfo
  {title} {{Magnetic Feshbach resonances in collisions of
  $^{23}\textrm{Na}^{40}\textrm{K}$ with $^{40}\textrm{K}$}},\ }\href
  {https://doi.org/10.1088/1367-2630/ac3318} {\bibfield  {journal} {\bibinfo
  {journal} {\textit{New J. Phys.}}\ }\textbf {\bibinfo {volume} {23}},\
  \bibinfo {pages} {115010} (\bibinfo {year} {2021})}\BibitemShut {NoStop}%
\bibitem [{\citenamefont {Morita}\ \emph {et~al.}(2024)\citenamefont {Morita},
  \citenamefont {Kosicki}, \citenamefont {Żuchowski}, \citenamefont {Brumer},\
  and\ \citenamefont {Tscherbul}}]{Morita2024}%
  \BibitemOpen
  \bibfield  {author} {\bibinfo {author} {\bibfnamefont {M.}~\bibnamefont
  {Morita}}, \bibinfo {author} {\bibfnamefont {M.~B.}\ \bibnamefont {Kosicki}},
  \bibinfo {author} {\bibfnamefont {P.~S.}\ \bibnamefont {Żuchowski}},
  \bibinfo {author} {\bibfnamefont {P.}~\bibnamefont {Brumer}},\ and\ \bibinfo
  {author} {\bibfnamefont {T.~V.}\ \bibnamefont {Tscherbul}},\ }\bibfield
  {title} {\bibinfo {title} {{Magnetic Feshbach resonances in ultracold
  atom-molecule collisions}},\ }\href
  {https://doi.org/10.1103/physreva.110.l021301} {\bibfield  {journal}
  {\bibinfo  {journal} {\textit{Phys. Rev. A}}\ }\textbf {\bibinfo {volume}
  {110}},\ \bibinfo {pages} {L021301} (\bibinfo {year} {2024})}\BibitemShut
  {NoStop}%
\bibitem [{\citenamefont {Hermsmeier}\ \emph {et~al.}(2021)\citenamefont
  {Hermsmeier}, \citenamefont {Kłos}, \citenamefont {Kotochigova},\ and\
  \citenamefont {Tscherbul}}]{Hermsmeier2021}%
  \BibitemOpen
  \bibfield  {author} {\bibinfo {author} {\bibfnamefont {R.}~\bibnamefont
  {Hermsmeier}}, \bibinfo {author} {\bibfnamefont {J.}~\bibnamefont {Kłos}},
  \bibinfo {author} {\bibfnamefont {S.}~\bibnamefont {Kotochigova}},\ and\
  \bibinfo {author} {\bibfnamefont {T.~V.}\ \bibnamefont {Tscherbul}},\
  }\bibfield  {title} {\bibinfo {title} {Quantum spin state selectivity and
  magnetic tuning of ultracold chemical reactions of triplet alkali-metal
  dimers with alkali-metal atoms},\ }\href
  {https://doi.org/10.1103/physrevlett.127.103402} {\bibfield  {journal}
  {\bibinfo  {journal} {\textit{Phys. Rev. Lett.}}\ }\textbf {\bibinfo {volume}
  {127}},\ \bibinfo {pages} {103402} (\bibinfo {year} {2021})}\BibitemShut
  {NoStop}%
\bibitem [{\citenamefont {Bird}\ \emph {et~al.}(2023)\citenamefont {Bird},
  \citenamefont {Tarbutt},\ and\ \citenamefont {Hutson}}]{RbandCaFMike2023}%
  \BibitemOpen
  \bibfield  {author} {\bibinfo {author} {\bibfnamefont {R.~C.}\ \bibnamefont
  {Bird}}, \bibinfo {author} {\bibfnamefont {M.~R.}\ \bibnamefont {Tarbutt}},\
  and\ \bibinfo {author} {\bibfnamefont {J.~M.}\ \bibnamefont {Hutson}},\
  }\bibfield  {title} {\bibinfo {title} {{Tunable Feshbach resonances in
  collisions of ultracold molecules in $^2\Sigma$ states with alkali-metal
  atoms}},\ }\href {https://doi.org/10.1103/physrevresearch.5.023184}
  {\bibfield  {journal} {\bibinfo  {journal} {\textit{Phys. Rev. Res.}}\
  }\textbf {\bibinfo {volume} {5}},\ \bibinfo {pages} {023184} (\bibinfo {year}
  {2023})}\BibitemShut {NoStop}%
\bibitem [{\citenamefont {Schwartz}\ \emph {et~al.}(2021)\citenamefont
  {Schwartz}, \citenamefont {Shimazaki}, \citenamefont {Kuhlenkamp},
  \citenamefont {Watanabe}, \citenamefont {Taniguchi}, \citenamefont {Kroner},\
  and\ \citenamefont {Imamoğlu}}]{Schwartz2021}%
  \BibitemOpen
  \bibfield  {author} {\bibinfo {author} {\bibfnamefont {I.}~\bibnamefont
  {Schwartz}}, \bibinfo {author} {\bibfnamefont {Y.}~\bibnamefont {Shimazaki}},
  \bibinfo {author} {\bibfnamefont {C.}~\bibnamefont {Kuhlenkamp}}, \bibinfo
  {author} {\bibfnamefont {K.}~\bibnamefont {Watanabe}}, \bibinfo {author}
  {\bibfnamefont {T.}~\bibnamefont {Taniguchi}}, \bibinfo {author}
  {\bibfnamefont {M.}~\bibnamefont {Kroner}},\ and\ \bibinfo {author}
  {\bibfnamefont {A.}~\bibnamefont {Imamoğlu}},\ }\bibfield  {title} {\bibinfo
  {title} {{Electrically tunable Feshbach resonances in twisted bilayer
  semiconductors}},\ }\href {https://doi.org/10.1126/science.abj3831}
  {\bibfield  {journal} {\bibinfo  {journal} {\textit{Science}}\ }\textbf
  {\bibinfo {volume} {374}},\ \bibinfo {pages} {336–340} (\bibinfo {year}
  {2021})}\BibitemShut {NoStop}%
\bibitem [{\citenamefont {Voges}\ \emph {et~al.}(2020)\citenamefont {Voges},
  \citenamefont {Gersema}, \citenamefont {Meyer~zum Alten~Borgloh},
  \citenamefont {Schulze}, \citenamefont {Hartmann}, \citenamefont {Zenesini},\
  and\ \citenamefont {Ospelkaus}}]{Kai}%
  \BibitemOpen
  \bibfield  {author} {\bibinfo {author} {\bibfnamefont {K.~K.}\ \bibnamefont
  {Voges}}, \bibinfo {author} {\bibfnamefont {P.}~\bibnamefont {Gersema}},
  \bibinfo {author} {\bibfnamefont {M.}~\bibnamefont {Meyer~zum
  Alten~Borgloh}}, \bibinfo {author} {\bibfnamefont {T.~A.}\ \bibnamefont
  {Schulze}}, \bibinfo {author} {\bibfnamefont {T.}~\bibnamefont {Hartmann}},
  \bibinfo {author} {\bibfnamefont {A.}~\bibnamefont {Zenesini}},\ and\
  \bibinfo {author} {\bibfnamefont {S.}~\bibnamefont {Ospelkaus}},\ }\bibfield
  {title} {\bibinfo {title} {{Ultracold Gas of Bosonic
  $^{23}\mathrm{Na}^{39}\mathrm{K}$ Ground-State Molecules}},\ }\href
  {https://doi.org/10.1103/PhysRevLett.125.083401} {\bibfield  {journal}
  {\bibinfo  {journal} {\textit{Phys. Rev. Lett.}}\ }\textbf {\bibinfo {volume}
  {125}},\ \bibinfo {pages} {083401} (\bibinfo {year} {2020})}\BibitemShut
  {NoStop}%
\bibitem [{\citenamefont {Shammout}\ \emph {et~al.}(2023)\citenamefont
  {Shammout}, \citenamefont {Karpa}, \citenamefont {Ospelkaus}, \citenamefont
  {Tiemann},\ and\ \citenamefont {Dulieu}}]{Shammout2023}%
  \BibitemOpen
  \bibfield  {author} {\bibinfo {author} {\bibfnamefont {B.}~\bibnamefont
  {Shammout}}, \bibinfo {author} {\bibfnamefont {L.}~\bibnamefont {Karpa}},
  \bibinfo {author} {\bibfnamefont {S.}~\bibnamefont {Ospelkaus}}, \bibinfo
  {author} {\bibfnamefont {E.}~\bibnamefont {Tiemann}},\ and\ \bibinfo {author}
  {\bibfnamefont {O.}~\bibnamefont {Dulieu}},\ }\bibfield  {title} {\bibinfo
  {title} {{Modeling Photoassociative Spectra of Ultracold NaK + K}},\ }\href
  {https://doi.org/10.1021/acs.jpca.3c01823} {\bibfield  {journal} {\bibinfo
  {journal} {\textit{J. Phys. Chem. A.}}\ }\textbf {\bibinfo {volume} {127}},\
  \bibinfo {pages} {7872–7883} (\bibinfo {year} {2023})}\BibitemShut
  {NoStop}%
\bibitem [{\citenamefont {Frye}\ and\ \citenamefont
  {Hutson}(2023)}]{SelectionRulesFrye}%
  \BibitemOpen
  \bibfield  {author} {\bibinfo {author} {\bibfnamefont {M.~D.}\ \bibnamefont
  {Frye}}\ and\ \bibinfo {author} {\bibfnamefont {J.~M.}\ \bibnamefont
  {Hutson}},\ }\bibfield  {title} {\bibinfo {title} {{Long-range states and
  Feshbach resonances in collisions between ultracold alkali-metal diatomic
  molecules and atoms}},\ }\href
  {https://doi.org/10.1103/PhysRevResearch.5.023001} {\bibfield  {journal}
  {\bibinfo  {journal} {\textit{Phys. Rev. Res.}}\ }\textbf {\bibinfo {volume}
  {5}},\ \bibinfo {pages} {023001} (\bibinfo {year} {2023})}\BibitemShut
  {NoStop}%
\bibitem [{\citenamefont {Shammout}\ \emph {et~al.}(2025)\citenamefont
  {Shammout}, \citenamefont {Karpa}, \citenamefont {Ospelkaus}, \citenamefont
  {Tiemann},\ and\ \citenamefont {Dulieu}}]{Shammout2025}%
  \BibitemOpen
  \bibfield  {author} {\bibinfo {author} {\bibfnamefont {B.}~\bibnamefont
  {Shammout}}, \bibinfo {author} {\bibfnamefont {L.}~\bibnamefont {Karpa}},
  \bibinfo {author} {\bibfnamefont {S.}~\bibnamefont {Ospelkaus}}, \bibinfo
  {author} {\bibfnamefont {E.}~\bibnamefont {Tiemann}},\ and\ \bibinfo {author}
  {\bibfnamefont {O.}~\bibnamefont {Dulieu}},\ }\bibfield  {title} {\bibinfo
  {title} {Formation of ultracold triatomic molecules by electric microwave
  association},\ }\href {https://doi.org/10.1103/physrevresearch.7.023187}
  {\bibfield  {journal} {\bibinfo  {journal} {\textit{Phys. Rev. Res.}}\
  }\textbf {\bibinfo {volume} {7}},\ \bibinfo {pages} {023187} (\bibinfo {year}
  {2025})}\BibitemShut {NoStop}%
\bibitem [{\citenamefont {Schulze}\ \emph {et~al.}(2018)\citenamefont
  {Schulze}, \citenamefont {Hartmann}, \citenamefont {Voges}, \citenamefont
  {Gempel}, \citenamefont {Tiemann}, \citenamefont {Zenesini},\ and\
  \citenamefont {Ospelkaus}}]{NaKmixture}%
  \BibitemOpen
  \bibfield  {author} {\bibinfo {author} {\bibfnamefont {T.~A.}\ \bibnamefont
  {Schulze}}, \bibinfo {author} {\bibfnamefont {T.}~\bibnamefont {Hartmann}},
  \bibinfo {author} {\bibfnamefont {K.~K.}\ \bibnamefont {Voges}}, \bibinfo
  {author} {\bibfnamefont {M.~W.}\ \bibnamefont {Gempel}}, \bibinfo {author}
  {\bibfnamefont {E.}~\bibnamefont {Tiemann}}, \bibinfo {author} {\bibfnamefont
  {A.}~\bibnamefont {Zenesini}},\ and\ \bibinfo {author} {\bibfnamefont
  {S.}~\bibnamefont {Ospelkaus}},\ }\bibfield  {title} {\bibinfo {title}
  {{Feshbach spectroscopy and dual-species Bose-Einstein condensation of
  $^{23}\mathrm{Na}\text{\ensuremath{-}}^{39}\mathrm{K}$ mixtures}},\ }\href
  {https://doi.org/10.1103/PhysRevA.97.023623} {\bibfield  {journal} {\bibinfo
  {journal} {\textit{Phys. Rev. A}}\ }\textbf {\bibinfo {volume} {97}},\
  \bibinfo {pages} {023623} (\bibinfo {year} {2018})}\BibitemShut {NoStop}%
\bibitem [{\citenamefont {Gempel}\ \emph {et~al.}(2016)\citenamefont {Gempel},
  \citenamefont {Hartmann}, \citenamefont {Schulze}, \citenamefont {Voges},
  \citenamefont {Zenesini},\ and\ \citenamefont {Ospelkaus}}]{Gempel2016}%
  \BibitemOpen
  \bibfield  {author} {\bibinfo {author} {\bibfnamefont {M.~W.}\ \bibnamefont
  {Gempel}}, \bibinfo {author} {\bibfnamefont {T.}~\bibnamefont {Hartmann}},
  \bibinfo {author} {\bibfnamefont {T.~A.}\ \bibnamefont {Schulze}}, \bibinfo
  {author} {\bibfnamefont {K.~K.}\ \bibnamefont {Voges}}, \bibinfo {author}
  {\bibfnamefont {A.}~\bibnamefont {Zenesini}},\ and\ \bibinfo {author}
  {\bibfnamefont {S.}~\bibnamefont {Ospelkaus}},\ }\bibfield  {title} {\bibinfo
  {title} {{Versatile electric fields for the manipulation of ultracold NaK
  molecules}},\ }\href {https://doi.org/10.1088/1367-2630/18/4/045017}
  {\bibfield  {journal} {\bibinfo  {journal} {\textit{New J. Phys.}}\ }\textbf
  {\bibinfo {volume} {18}},\ \bibinfo {pages} {045017} (\bibinfo {year}
  {2016})}\BibitemShut {NoStop}%
\bibitem [{\citenamefont {Gerdes}\ \emph {et~al.}(2011)\citenamefont {Gerdes},
  \citenamefont {Dulieu}, \citenamefont {Kn\"{o}ckel},\ and\ \citenamefont
  {Tiemann}}]{Gerdes2011}%
  \BibitemOpen
  \bibfield  {author} {\bibinfo {author} {\bibfnamefont {A.}~\bibnamefont
  {Gerdes}}, \bibinfo {author} {\bibfnamefont {O.}~\bibnamefont {Dulieu}},
  \bibinfo {author} {\bibfnamefont {H.}~\bibnamefont {Kn\"{o}ckel}},\ and\
  \bibinfo {author} {\bibfnamefont {E.}~\bibnamefont {Tiemann}},\ }\bibfield
  {title} {\bibinfo {title} {{Stark effect measurements on the NaK molecule}},\
  }\href {https://doi.org/10.1140/epjd/e2011-20048-9} {\bibfield  {journal}
  {\bibinfo  {journal} {\textit{Eur. Phys. J. D.}}\ }\textbf {\bibinfo {volume}
  {65}},\ \bibinfo {pages} {105–111} (\bibinfo {year} {2011})}\BibitemShut
  {NoStop}%
\end{thebibliography}%

\end{document}